\documentclass[trackchanges]{aastex631}

\begin{document}

\title{Applied Machine-Learning Models to Identify Spectral Sub-Types of M Dwarfs \\ from Photometric Surveys}

\author[0000-0002-3392-6956]{Sirinrat Sithajan}
\author[0009-0006-2813-7799]{Sukanya Meethong}
\affiliation{National Astronomical Research Institute of Thailand \\
Don Kaeo, Mae Rim, Chiang Mai, 50180, Thailand}




\begin{abstract}

M dwarfs are the most abundant stars in the Solar Neighborhood and they are prime targets for searching for rocky planets in habitable zones. Consequently, a detailed characterization of these stars is in demand. The spectral sub-type is one of the parameters that is used for the characterization and it is traditionally derived from the observed spectra. However, obtaining the spectra of M dwarfs is expensive in terms of observation time and resources due to their intrinsic faintness. We study the performance of four machine-learning (ML) models—K-Nearest Neighbor (KNN), Random Forest (RF), Probabilistic Random Forest (PRF), and Multilayer Perceptron (MLP)—in identifying the spectral sub-types of M dwarfs at a grand scale by deploying broadband photometry in the optical and near-infrared. We trained the ML models by using the spectroscopically identified M dwarfs from the Sloan Digital Sky Survey (SDSS) Data Release (DR) 7, together with their photometric colors that were derived from the SDSS, Two-Micron All-Sky Survey (2MASS), and Wide-field Infrared Survey Explorer (WISE). We found that the RF, PRF, and MLP give a comparable prediction accuracy, 74\%, while the KNN provides slightly lower accuracy, 71\%. We also found that these models can predict the spectral sub-type of M dwarfs with $\sim$99\% accuracy within ±1 sub-type. The five most useful features for the prediction are \textit{r-z}, \textit{r-i}, \textit{r-J}, \textit{r-H}, and \textit{g-z}, and hence lacking data in all SDSS bands substantially reduces the prediction accuracy. However, we can achieve an accuracy of over 70\% when the \textit{r} and \textit{i} magnitudes are available. Since the stars in this study are nearby ($d\lesssim$1300 pc for 95\% of the stars), the dust extinction can reduce the prediction accuracy by only 3\%. Finally, we used our optimized RF models to predict the spectral sub-types of M dwarfs from the Catalog of Cool Dwarf Targets for the Transiting Exoplanet Survey Satellite (TESS), and we provide the optimized RF models for public use.

\end{abstract}

\keywords{}


\section{Introduction} \label{sec:intro}

M dwarfs are the major population in the Galaxy \citep{Henry2006, Bochanski2010} and they have received attention as potential targets for detecting terrestrial-mass planets in habitable zones \citep{Alonso2015, Muirhead2018}. The difficulty in characterizing M dwarfs, especially when their properties are traditionally derived from spectroscopy (e.g., the accurate spectral type, surface temperature and gravity, and chemical abundance), is their intrinsic faintness. A number of studies, though not limited to M dwarfs, have tried to use the data from photometric surveys to calibrate spectroscopic stellar properties because photometric data are relatively easier and faster to obtain than spectroscopic data. For example, \citet{Mann2015, Mann2016} derived an empirical relation between stellar effective temperature (\textit{T}\textsubscript{eff}) and broadband photometry for cool stars, and \citet{Mucciarelli2021} presented an estimation of \textit{T}\textsubscript{eff} from Gaia \citep{Gaia2016} photometry. 

While these studies derived a stellar property empirically, some studies have used machine-learning techniques, which recently become popular in analyzing astronomical data (see \citealt{Baron2019} and the references therein). \citet{Miller2015} used machine learning to estimate [Fe/H] to discover extremely metal-poor (EMP) stars from broadband photometry and received a [Fe/H] as good as that derived from low-resolution (R$\sim$1800) spectroscopy. \citet{Hardegree2019} used a machine-learning technique, called the Random Forest (RF) classifier, $r$, $J$ and $K_s$ magnitudes to classify the M dwarfs that were observed by Kepler \citep{Borucki2010} and were able to classify them more than $\sim$90\% accurately within one spectral sub-type of the sub-type obtained from the spectroscopic technique. \citet{Gong2022RNAAS} and \citet{Gutierrez2022} also used the RF and photometric colors to distinguish ultracool dwarfs from background contaminants and to classify them into spectral subgroups. \citet{Hardegree2020} employed the RF classifier and RF regression to derive the spectral type, \textit{T}\textsubscript{eff}, surface gravity (log \textit{g}) and [Fe/H] for $\sim$200000 \textit{K2} stars of mid-A to mid-M type that were lacking of the spectroscopically derived parameters.

Machine learning is a branch of computer science and artificial intelligence (AI), which is broadly defined as the use of data and algorithms to imitate human intelligence, behavior, and the way that humans learn. The recent increasing availability of computing resources and large data sets, as well as the improvement in the performance and usability of algorithms implemented in machine-learning software, have led to the success of its applications in various fields, including astronomy. Specifically, supervised learning, which is an algorithm that learns from known examples where both inputs and outputs are provided, can generalize to create a predicted output for an input that it has never seen before. When it is applied to problems where the output is a category, such as stellar spectral type, it is called a classification model or a classifier. 

Stellar spectral type is one of the fundamental parameters for characterizing stars. It groups together, to some degree, stars that have similar properties. Originally, the spectral type was assigned to a star based purely on the appearance of the stellar spectrum (see \citealt{Gray2009} and reference therein). The connection between the spectral appearance and some intrinsic stellar properties, such as \textit{T}\textsubscript{eff}, was found later. A number of studies have assessed the relationship between the stellar spectral type and other physical properties, such as the planet occurrence rate (e.g. \citealt{Fressin2013, Burke2015, Hardegree2019}).

Broadband photometric magnitude, or brightness, can be considered as a value that represents a part of stellar spectrum. The magnitudes from multiple wave bands can be thought of as an extremely low-resolution spectrum with the wavelength coverage of these wave bands. Given that the shapes of spectral energy distribution (SED) of stars with different spectral types are different, the relative values of the magnitudes from given pairs of wave bands of these stars will also be different. Therefore, photometric color, which is defined as the brightness ratio of a pair of wave bands, can be used as a feature for machine learning. By providing the colors of stars as the input (features) and the corresponding spectral types as the known output (labels) for machine learning, a classification model can be created and it can be used to predict the spectral type of a star given only input colors. 

This study aims to investigate the ability of machine-learning (ML) algorithms to classify M dwarfs into spectral sub-type (M0V-M9V) by using only photometric information. The four different algorithms that we consider in this study are K-Nearest Neighbor (KNN), RF, Probabilistic Random Forest (PRF), and Multilayer Perceptron (MLP). Each of these algorithms has its own characteristics that could make it a useful model for our problem. KNN is one of the simplest ML algorithms and can be used as a baseline for our study. RF is relatively fast, interpretable, and has been widely used successfully in many applications. PRF resembles RF but is able to natively handle missing data, and take into account the uncertainty of the features and the labels. MLP, in general, has the ability to learn complex nonlinear pattern, although in this study we only explore a few basic configurations of MLP. The second purpose of this study is to find the best ML model among the four studied algorithms and to identify the most useful wavelength regions defined by well-known broadband filters in the optical and Near-Infrared (NIR) for the classification. It is important to note that when building ML models in this study, we use the labels (i.e., the spectral sub-types) that are derived from optical spectra (specifically, $\Delta\lambda$$\sim$3800 - 9300$\AA$). For an M dwarf, the spectral sub-type derived from NIR spectra can be slightly different from that derived from the optical counterparts (e.g. \citealt{Hardegree2020}). However, optical spectra have been historically used to derive spectral types of M dwarfs and the other early- type stars \citep{Gray2009}. Therefore, the spectral types obtained from optical spectra are widely used and consistent among stars of all classes.    

In this work, we assume that the stars selected for constructing our ML models are dwarfs and are single stars, although the main catalog from which we select the stars (i.e., \citealt{West2011}) has 0.5\% contamination and \citet{Winters2019} estimated that around one-third of M dwarfs in the Solar Neighborhood possess stellar companions. The additional steps that are described in the later sections have been performed to solidify these assumptions. The rest of this paper is structured as follows. Section~\ref{sec:Data} describes the data that we used to create the classification models. Section~\ref{sec:Method} explains the ML methods. The results are shown in Section~\ref{sec:results}. In Section~\ref{sec:app}, we use our best ML models to classify the spectral sub-types of M dwarfs in the Catalog of Cool Dwarf Targets for TESS  \citep{Muirhead2018}. We describe the models that we release for public use and we explain how to use them in Section~\ref{sec:models}. A summary of this study is given in Section~\ref{sec:summary}.

\section{Data}
\label{sec:Data}
\citet{West2011}, hereafter W11, provide a spectroscopic catalog of M dwarfs from the SDSS \citep{York2000} DR7 \citep{Abazajian2009}. The catalog lists both broadband photometry and spectroscopic properties, including spectral sub-type, for 70,840 stars. Therefore, it is an ideal data set for constructing ML models. The available photometry includes SDSS \textit{ugriz} magnitudes and 2MASS \textit{JHK\textsubscript{s}} magnitudes. For the SDSS magnitudes, the catalog also provides dust extinction corrections, which were calculated based on \citet{Schlegel1998} 2D dust maps. The 2MASS magnitudes were obtained from the 2MASS point-source catalog \citep{Cutri2003} by cross-matching the stars in the W11 catalog with the positions in the 2MASS catalog that lie within 5$\arcsec$. Only the uniquely matched stars with high-quality 2MASS photometry were incorporated into the W11 catalog. The spectral sub-types, M0V-M9V, were derived by visual inspecting each observed SDSS spectrum (R$\sim$1800,  $\Delta\lambda$$\sim$3800 - 9300$\AA$) with the Hammer spectral typing facility \citep{Covey2007}. The estimated dispersion of the classified spectral types is $\sim$0.4 sub-type, which we will adopt as the error of spectral type when the ML algorithms require the value. 

Given that the shape of the stellar spectrum and the relative brightness from a given pair of photometric bands are unique for each spectral type, we include W1 and W2 magnitudes from AllWISE catalog \citep{Cutri2014}, which cover $\sim$3-5 $\mu$m, to increase the number of features for training the ML model and to see if they can help improve our classification models. W3 and W4 are not included because, for the majority of them, only the upper limit values of the brightness are available.
We use the CDS X-Match service\footnote{\label{foot:cdsx}\url{http://cdsxmatch.u-strasbg.fr/}} to map the SDSS positions in the W11 catalog to the positions of stars in the AllWISE catalog that fall within 3$\arcsec$. Although we adopt the 3$\arcsec$ angular separation, more than 94\% and 98\% of the matched stars fall within 1.0$\arcsec$ and 1.5$\arcsec$, respectively. Because 2MASS \textit{JHK\textsubscript{s}} magnitudes are available in both W11 and AllWISE catalogs, we compare these values of the matched stars. We see that they are all consistent, and so we incorporate the AllWISE magnitudes of the matched stars to the W11 catalog. 

\begin{figure}
	\includegraphics[width=1.0\textwidth]{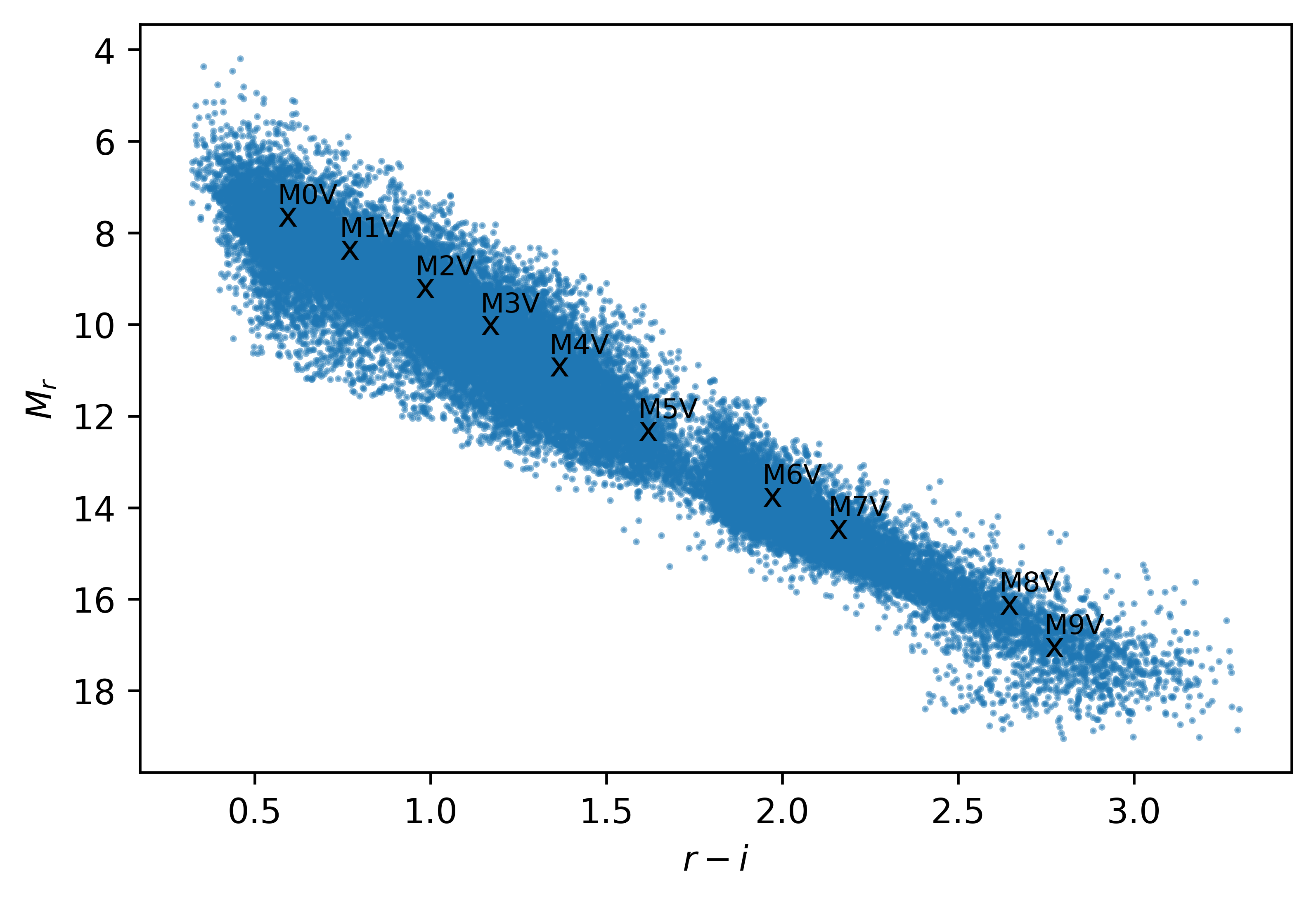}
    \caption{Color-magnitude diagram of the selected stars described in Section~\ref{sec:Data}. \textit{Mr} is the absolute magnitude in the SDSS-\textit{r} band calculated using the geometric distance from ~\citet{2021Bailer} and the 3D dust map of ~\citet{Green2018}. Note that \textit{r-i} is the dust-corrected \textit{r-i} color. The cross symbol indicates the location of the median value of the stars in each spectral sub-type on the diagram.}
    \label{fig:cmd}
\end{figure}

We select stars for building ML models when the stars satisfy all of the following conditions: (1) magnitudes and the corresponding errors are available in all \textit{ugrizJHK\textsubscript{s}}W1W2 bands, (2) have geometric distances from \citet{2021Bailer}, (3) possess the flags in W11 catalog indicating good photometry (\texttt{q} = 1) and  not white-dwarf–M-dwarf pair (\texttt{WDM} = 0), (4) have \texttt{non-single-star} flag = 0 in the Gaia DR3 catalog \citep{Gaia2022}, and (5) not in the \citet{El-Badry2021} catalog of binaries from Gaia eDR3. Although the W11 catalog indicates that its giant contamination rate is only 0.5\%, we apply an additional step to enhance the purity of our sample by removing clear outliers on the \textit{r-i} versus \textit{Mr} plane.
Finally, we have 39,676 stars that satisfy the conditions and we use these stars for creating the ML models. The statistical properties of these stars and their distribution on \textit{r-i} versus \textit{Mr} plane are shown in Table~\ref{tab:samplestat} and Figure~\ref{fig:cmd}.

\begin{table}
 \caption{Statistical properties of the stars used to train and evaluate ML models. The distances are geometric distances from ~\citet{2021Bailer}. \textit{Mr} is the absolute magnitude in the SDSS-\textit{r} band calculated using the geometric distance and the 3D dust map of ~\citet{Green2018}. \textit{r-i} is the dust-corrected \textit{r-i} color.}
 \label{tab:samplestat}
 \begin{tabular}{ccccccc}
  \hline
  Sub-Type & Counts & Median \textit{Mr} & Median \textit{r-i} & &  Distance (pc) at the Percentile & \\
  & & & & 16th & 50th & 84th \\
  \hline
  0 & 4850 & 7.86 & 0.565 & 519 & 861 & 1380 \\ 
  1 & 4005 & 8.58 & 0.741 & 433 & 724 & 1152 \\ 
  2 & 5473 & 9.42 & 0.956 & 325 & 516 & 842 \\ 
  3 & 6204 & 10.23 & 1.142 & 243 & 397 & 672 \\
  4 & 5001 & 11.13 & 1.338 & 176 & 314 & 585 \\
  5 & 2399 & 12.54 & 1.590 & 129 & 231 & 423 \\
  6 & 4601 & 13.98 & 1.943 & 129 & 201 & 291 \\
  7 & 5110 & 14.68 & 2.131 & 114 & 176 & 256 \\
  8 & 1358 & 16.33 & 2.617 & 82 & 121 & 175 \\
  9 & 675 & 17.26 & 2.745 & 63 & 96 & 150 \\
  \hline
  All & 39676 & 10.63 & 1.218 & 158 & 348 & 806 \\
  \hline
 \end{tabular}
\end{table}

\section{Methods}
\label{sec:Method}
\subsection{Training the Machine Learning Model}
\label{sec:training} 

In supervised learning, an ML algorithm is fed with examples, each consisting of a pair of an input and a corresponding output. The algorithm then learns the pattern and finds a way to predict an output given a new input. The learning process is called \textit{training}. The input could be a collection of properties describing the example, known as independent variables or \textit{features}. The output is usually called the dependent variable or \textit{label}. In this study, the features are constructed from the photometric information of the stars and the labels are the spectral sub-types of those stars.

The constructed models are expected to \textit{generalize} to be able to predict the spectral sub-type of stars with only photometric information provided. Therefore, a subset of data needs to be excluded from the learning process and only used to evaluate the model’s performance. The subset of data used in the training is called a \textit{training set} and the one that is used for the evaluation is called a \textit{test set}. We split the data set described in Section~\ref{sec:Data} by stratified sampling, keeping 80\% of the data for the training set and the remaining 20\% for the test set.

For each star, we construct 45 photometric colors from the dust-corrected magnitudes in 10 bands,  \textit{ugrizJHK\textsubscript{s}}W1W2. For example, color \textit{u-r} is obtained by the dust corrected \textit{u}-band magnitude minus the dust corrected \textit{r}-band magnitude. We calculate the dust extinction (\textit{A}) in each band using the 3D dust map of \citet{Green2018} and the geometric distance of \citet{2021Bailer}. Since \citet{Green2018} do not provide the extinction vector (\textit{R}) in the \textit{u}, W1, and W2 bands, we extrapolate the values from the other available bands. The corresponding error of each color is calculated using the standard error propagation formula. These photometric colors and the errors are used as features for the ML algorithms, which will be described in later sections.

Some ML algorithms, such as KNN and MLP, are sensitive to the scales of the features, so they need to be adjusted to have a similar scale. We explore two feature scaling approaches in this study. The first approach, called standardization, scales each feature so that the mean is 0 and the standard deviation is 1. The second approach, called min-max scaling, scales each feature so that the minimum value is 0 and the maximum value is 1. 

When one or more classes have very low proportions when compared to the other classes, it is referred to as an imbalanced classification problem. Since most of the classification algorithms assume equally distributed classes, a severely imbalanced training set can result in a model with poorly predictive performance, especially for the minority classes (most of the contemporary works on class imbalance concentrate on imbalance ratios ranging from 1:4 up to 1:100; \citealt{Krawczyk2016}). In most of the use cases, the minority class is the one that is important, and hence special techniques are needed to be applied to improve the performance of the model on the minority class. In our sample the ratios of the most minor classes M9V, M8V, and M5V to the most abundant class M3V are 1:9, 1:5, and 1:3, respectively (Table~\ref{tab:samplestat}), and all classes are equally important. We investigate two data sampling techniques that are commonly used to mitigate the imbalanced data, (1) Random Undersampling (RUS) and (2) Synthetic Minority Oversampling Technique (SMOTE, \citealt{Chawla2011}), to see how they affect the overall prediction and the prediction of each class. The library \textit{imblearn}\footnote{\label{foot:imblearn}\url{https://imbalanced-learn.org}} has these techniques implemented and the default settings are used for our study.

\subsection{Machine-Learning Algorithm}
\label{sec:MLalgorithm} 

In this study, we consider four different machine-learning algorithms, of which a brief overview is provided in the following subsections.

\subsubsection{K-nearest Neighbors}
\label{sec:KNN} 

KNN is one of the simplest machine-learning algorithms. In multi-class classification, for an unlabeled example, KNN identifies the k closest training examples in the feature space and counts how many of them belong to each class. The prediction for the unlabeled example will be the most common class. The KNN has two important hyperparameters-- the number of neighbors (k) and the way the distances between the examples are measured. In this study, we use Euclidean distance which is the simplest and appropriate because all of our features are numerical variables. We use the KNN implemented in the library \textit{scikit-learn}\footnote{\label{foot:sklearn}\url{https://scikit-learn.org}}.

\subsubsection{Random Forest}
\label{sec:RF}

The RF algorithm utilizes a technique called ensemble learning, in which a final prediction is made by aggregating the predictions of a group of classifiers. For RF, these classifiers are Decision Tree classifiers, and the final prediction is the class that obtains the most votes. Each of these decision trees is constructed using a different random subset of the training set (the bagging technique). When growing each tree, only a random subset of features is used to split a node. The randomness introduced by the bagging and random features reduces the variance of the overall model relative to using single Decision Tree classifiers and generally yields better performance. Some of the hyperparameters of RF are inherited from Decision Tree classifier, such as the maximum depth of each decision tree (max\_depth) and the number of features used to split a node (max\_features). The other hyperparameters are for controlling the ensemble, such as the total number of decision trees constituting the forest (n\_estimators). In this study, we use the RF implemented in \textit{scikit-learn}.

\subsubsection{Probabilistic Random Forest}
\label{sec:PRF}
Probabilistic Random Forest\footnote{\label{foot:prf}\url{https://github.com/ireis/PRF}} (PRF; \citealt{Reis2019}) is a modified and improved version of the RF algorithm that includes the measurement uncertainty of the features or that of the label. In addition, PRF is able to handle missing data in the data set. For the uncertainty in the features, while the original RF assumes no uncertainty and an instance of data in a given node propagates either to the subsequent right or left node, the PRF uses the cumulative probability distribution function (PDF) for a particular feature to determine the probability of propagating to the left or right branch. Therefore, an instance can pass to both branches (with some probability). In principle, any function can be used for the PDF. However, in our PRF implementation, the Gaussian function is used. The node split is guided by the modified version of Gini impurity, which takes into account the uncertainty in the label. Once a tree is built, each leaf node would have a class probability. When making the prediction for a new instance, for each class, PRF sums the class probability from all leaf nodes weighted by the probability of that instance being in each leaf node.

\subsubsection{Multilayer Perceptron}
\label{sec:MLP}
MLP is a class of the Artificial Neural Network (ANN) algorithms, where the architecture consists of one or more hidden layers. In this study, the MLP has the first hidden layer with 300 neurons and one or two additional hidden layers with 100 neurons each. Every layer is fully connected (dense). For the hidden layers, the activation function is the Rectified Linear Unit Function (ReLu). For the output layer, the activation function is softmax activation. Regarding the loss function, cross-entropy loss is used. When training, the Stochastic Gradient Descent (SGD) optimizer is used. Our hyperparameters to be optimized are the learning rate and the number of additional hidden layers (i.e., 1 or 2). Regarding regularization, a dropout of 20\% is applied on hidden layers. In this study, we use the MLP implemented in \textit{Keras}\footnote{\label{foot:keras}\url{https://keras.io}}.

\subsection{Parameter Tuning and Evaluation Metric}
\label{sec:GCV}

Most ML algorithms have parameters that have to be specified before learning from the data. These parameters are called the model's hyperparameters. Some important hyperparameters for each algorithm are described in Section~\ref{sec:MLalgorithm}. We find the optimized set of hyperparameters (parameter tuning) using 10-fold cross-validation and a grid search over the hyperparameter space. The option of performing scaling or resampling and which type should be used are treated as one of the hyperparameters in the tuning process. The best set of hyperparameters is selected based on the optimal accuracy, which is defined as the ratio of the correctly predicted examples to the total examples, and is used for training the final ML model. The final ML model is used to make predictions for the examples in the test set. The accuracy of the final model on the test set is used as a performance metric when we report our results and when we make a comparison of the performance of the different models.     

\section{Results}
\label{sec:results}

\subsection{Accuracy of the Machine-Learning Models}
\label{sec:bestml}

\begin{table}
 \caption{Accuracy of the optimized ML models built from all 45 features. The training time and prediction time are the CPU time (a laptop with a 2.3 GHz Intel Processor) used in training a model and making predictions of the stars in the test set, respectively. 
 }
 \label{tab:acc_result}
 \begin{tabular}{ccccc}
  \hline
    Model & Accuracy & Accuracy & Training Time & Prediction Time \\
     & (Exact Sub-Type) & (Including $\pm$1 Sub-Type) &  &  \\
  \hline
  KNN & 0.712 $\pm$ 0.002 & 0.993 & 0.006 s & 6 s   \\
  RF  & 0.738 $\pm$ 0.003 & 0.996 & 1 minutes 37 s & 0.3 s   \\
  PRF & 0.741 $\pm$ 0.003 & 0.996 & 44 minutes 35 s & 47 s  \\
  MLP & 0.737 $\pm$ 0.004 &  0.995 & 5 minutes 30 s & 0.6 s  \\
  \hline
 \end{tabular}
\end{table}

\begin{figure}
	\includegraphics[width=1.0\textwidth]{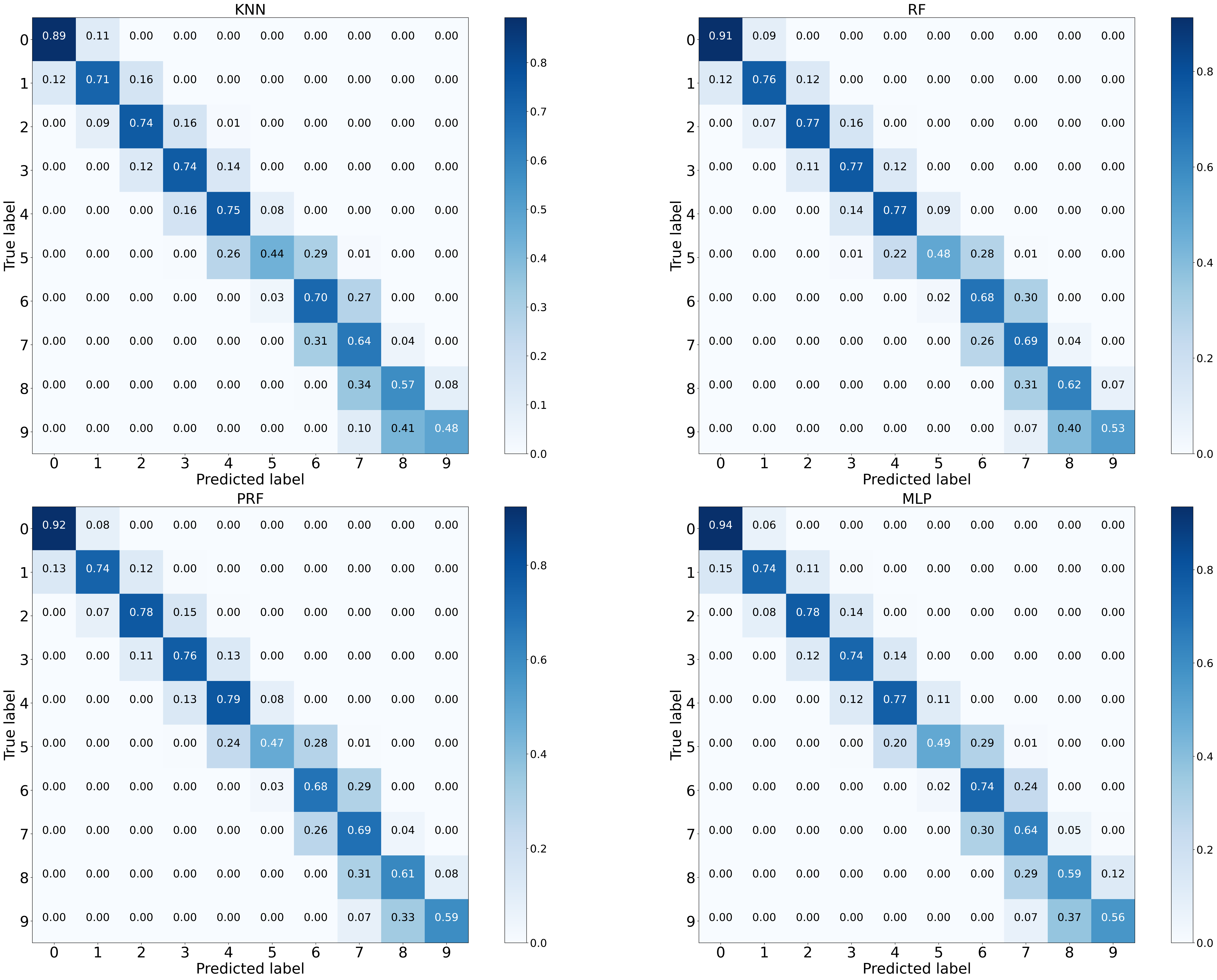}
    \caption{Normalized confusion matrices of the optimized ML models built using 45 features. Each matrix is normalized such that each row has the summation of one. The main diagonal cells indicate the correct predictions of the model for given spectral sub-types.}
    \label{fig:cfm}
\end{figure}

\begin{figure}
	\includegraphics[width=0.9\textwidth]{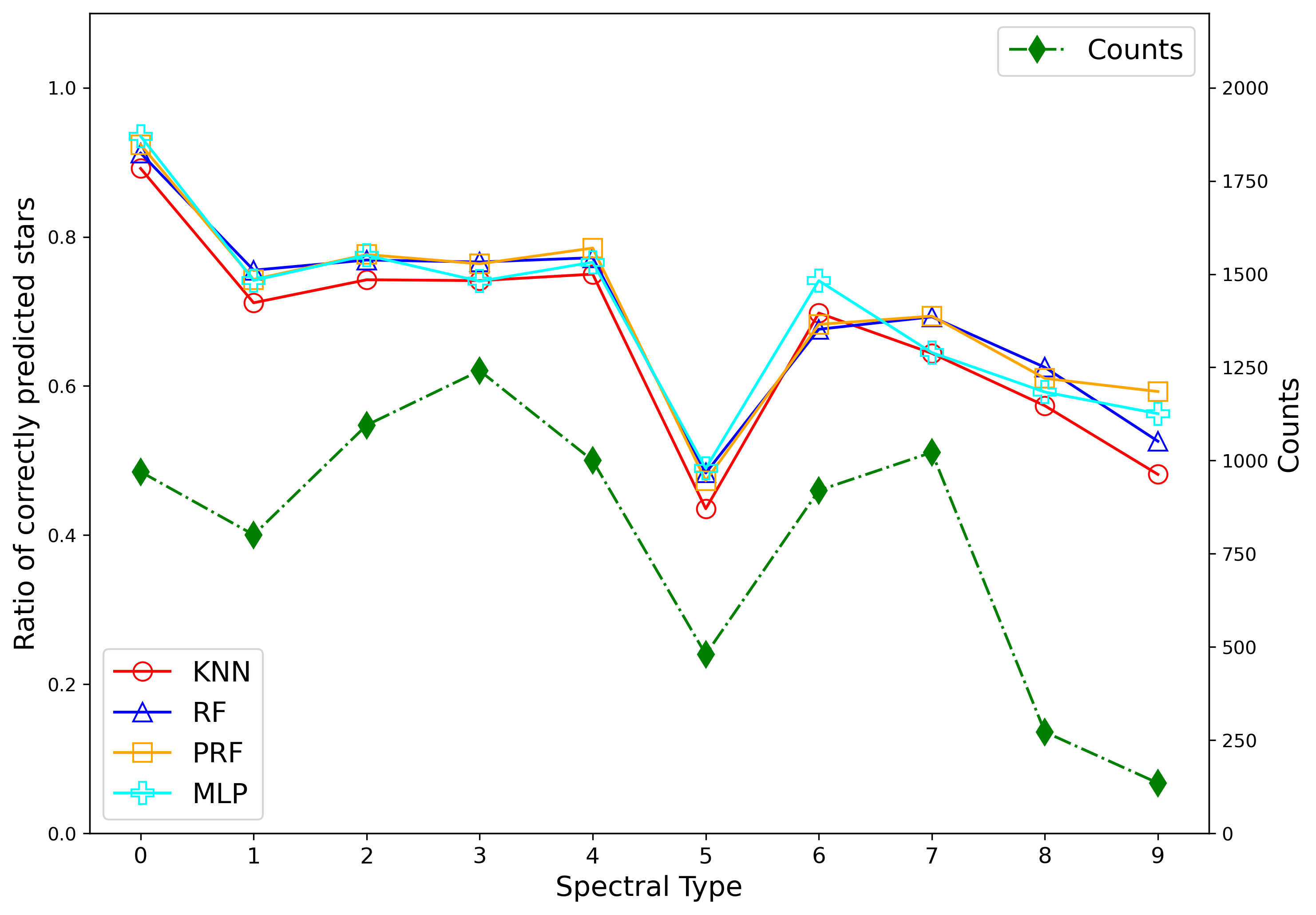}
    \caption{The ratio of correct prediction, or the value in the main diagonal cell in  Figure~\ref{fig:cfm}, (solid line) and the number of stars (dashed line) vs. spectral sub-type. Different colors illustrate the results for different ML methods, including KNN, RF, PRF, and MLP. The lines are drawn only for clarity and no information is contained between the data points.}
    \label{fig:recallspec}
\end{figure}

We compare the accuracy of the four models built from all 45 features and the optimized sets of hyperparameters. We find that the RF, PRF, and MLP models give a comparable overall prediction accuracy, which is $\sim$74\%, and the KNN provides slightly lower accuracy, which is $\sim$71\%. The accuracy of these models and the corresponding uncertainty are shown in Table~\ref{tab:acc_result}. The uncertainties are the standard error of the mean test score obtained from the cross-validation. Figure~\ref{fig:cfm} shows the confusion matrices of these models. For each confusion matrix, each main diagonal cell displays the fraction of correct prediction, while its left-hand and right-hand cells show the fractions of wrong predictions; that is, the matrix is normalized such that each row has the summation of one. Figure~\ref{fig:recallspec} shows plots of the value in the main diagonal cell, or the fraction of correct predictions, versus spectral sub-type (solid lines). We see that all models yield a similar trend. When comparing the plots to the number of stars in each spectral sub-type (dashed line), we can see some correlation (i.e., the fraction of correct prediction decreases when the number of stars used in training the ML decreases). However, they are not directly proportional. This means there are other factors beside the number of stars that affect the prediction accuracy.

Better performance is expected when applying feature scaling to the KNN. However, in our case, both min-max scaling and standardization do not improve the overall accuracy. For the MLP, the standardization gives the best accuracy. Theoretically, feature scaling would have no effect for RF and PRF, and our results confirm this. 

\begin{figure}
	\includegraphics[width=0.9\textwidth]{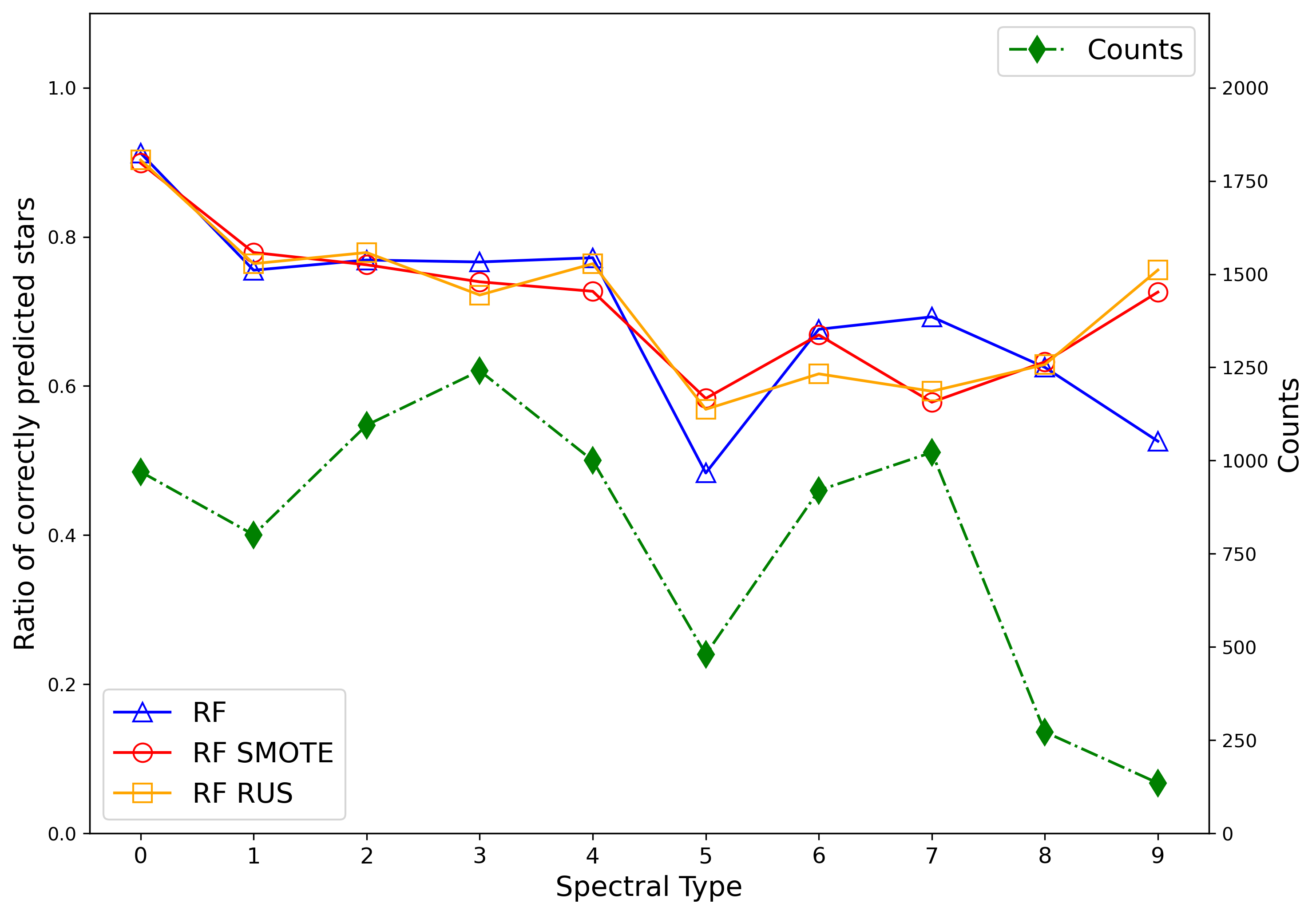}
    \caption{Comparison of the ratio of correctly predicted stars in each spectral sub-type for the RF models trained with and without resampling (solid line). The considered resampling techniques are Synthetic Minority Oversampling Technique (SMOTE) and Random Undersampling (RUS). The number of stars in each spectral sub-type is also shown (dashed line). The lines are drawn only for clarity and no information is contained between the data points.}
    \label{fig:recallspec_resamp}
\end{figure}

When either SMOTE or RUS is applied in the training, a significant improvement in the minority class prediction (M5V, M8V, and M9V) can be seen in all models. For the RF model, which gives the best overall accuracy among the models with the same resampling technique, the ratio of the correctly predicted stars increases by 21\%, 1.2\%, and 38\% when SMOTE is applied and increases by 18\%, 0.6\%, and 44\% when RUS is applied, for M5V, M8V, and M9V sub-type, respectively. However, the percentage of the increase for the sub-type does not depend on its ratio to the most abundance class. The effect of applying resampling is illustrated in Figure~\ref{fig:recallspec_resamp}.
Although there is an improvement in the minority classes predictions, with our test set that has an uneven sub-type distribution inherited from the SDSS survey (as displayed in Table~\ref{tab:samplestat}), applying SMOTE or RUS results in a 2\% decrease of the overall accuracy. We further found that, if the sub-types of the examples in the test set were uniformly distributed, the RF models with resampling would give an overall accuracy higher than the model without resampling by 3\%. In other words, if the distribution of spectral sub-types in the sample to be predicted is different from that in the training sample, the overall prediction accuracy will slightly decrease. In practice, it is unlikely for there to be prior knowledge of the distribution of the sub-types in the sample to be classified, and hence a slight decrease in the prediction accuracy should be expected. In this work, when we train the final models, which are selected based on the overall accuracy as described in ~\ref{sec:GCV}, we do not apply any resampling unless we state this explicitly.

\begin{figure*}
	\includegraphics[width=1.0\textwidth]{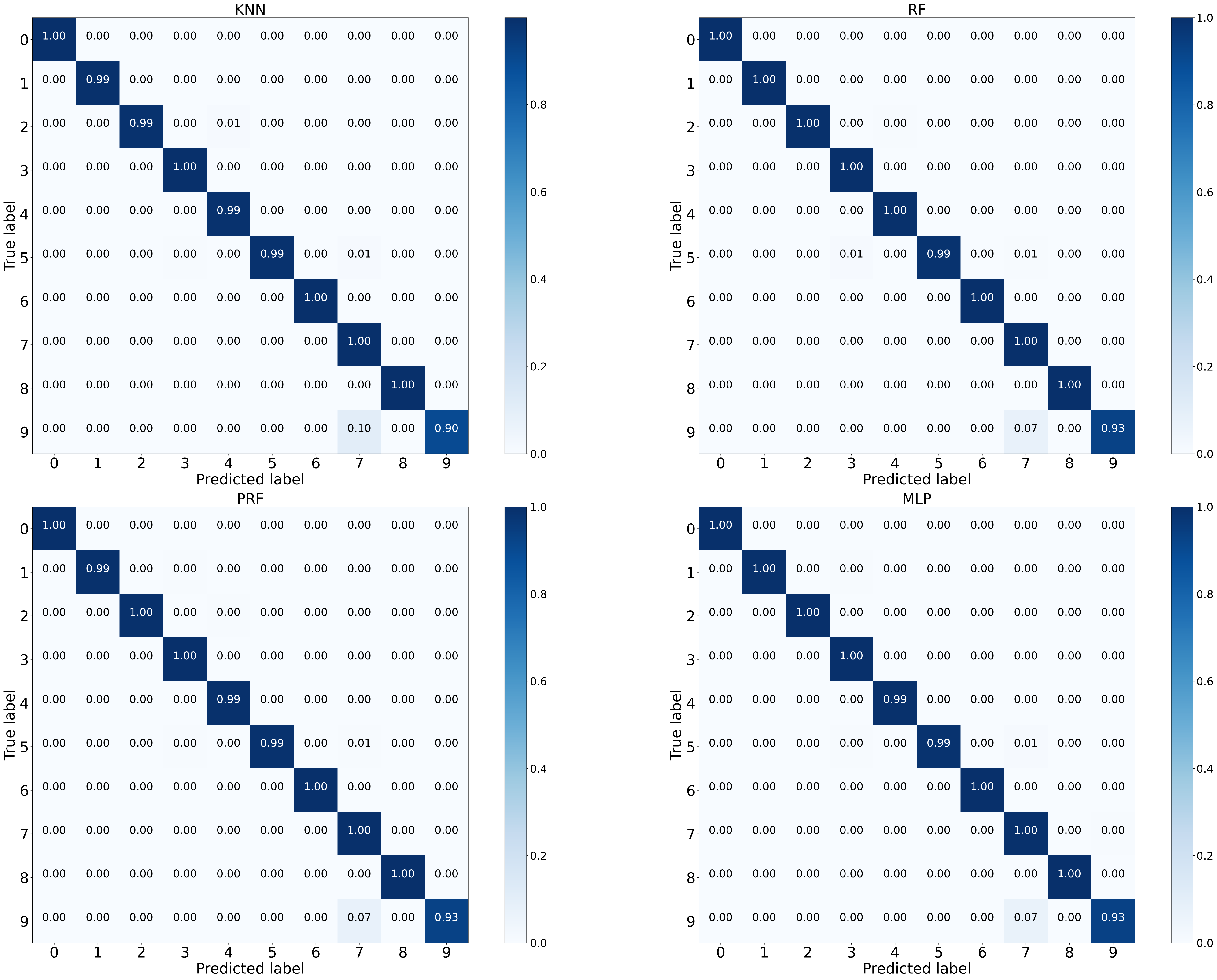}
    \caption{Normalized confusion matrices of the optimized ML models built using 45 features. Each matrix is normalized such that each row has the summation of one. The main diagonal cells indicate the correct predictions of the model for given spectral sub-types when allowing for the error of $\pm$1 sub-type.}
    
    \label{fig:cfm1}
\end{figure*}

Generally, the finest spectral sub-type grid considered when studying a trend of astrophysical properties is 1 sub-type. We estimate the accuracy of the best models when allowing the error of $\pm$1 sub-type of a given sub-type and the result is shown in Table~\ref{tab:acc_result} and Figure~\ref{fig:cfm1}. We see that the prediction is 99-100\% correct within $\pm$1 sub-type for most of the spectral sub-types, and the overall accuracy is 99-100\%. These results indicate that by using only broadband magnitudes in the optical and NIR and a basic ML technique, we can estimate the spectral sub-type of M dwarfs with $\sim$99-100\% accuracy within 1 sub-type.

Although the RF and PRF models give a comparable result, the RF model could be built much more quickly. The PRF algorithm took into account the measurement uncertainty, including photometric error and spectral type error, explicitly when it trained the model. However, we do not see significant improvement in the accuracy when compared to the RF model. This may happen because the RF can also capture the uncertainty in the data, specifically this data set, by itself. As mentioned in Section~\ref{sec:PRF}, the PRF has the ability to handle missing data. We do not consider this here, but defer to Section~\ref{sec:missingfeature}.

\subsection{Missing Features}
\label{sec:missingfeature}

In practice, the measured magnitudes are not always available in all 10 bands, like those used in training the models. We study how the missing measurements affects the accuracy of the models by investigating the following scenarios. First, we assume that we have the measurements from all 10 bands, but a fraction of stars have some missing values and the missing values are random. In this scenario, we can compare the performance of the PRF in handling missing values to the performance of the RF where the missing values are imputed using either mean imputation or KNN imputation. It is necessary that the features of the example to be predicted are the same as the features used to train the model. Therefore, imputation of missing features is required before making a prediction. We randomly nullify a portion of the data in the test set, as follows. We perform four experiments by varying the amount of missing data. For each feature, we randomly nullify 5\%, 10\%, 15\%, and 50\% of the data. For PRF, the missing data are handled natively. For the RF, using the optimized RF model built previously, we impute those missing data by either mean imputation or KNN imputation (with k = 20). We obtain the prediction accuracy that is shown in Table~\ref{tab:missfeatures}. We see that when the fraction of missing data is small, the mean imputation and KNN imputation yield a comparable accuracy. However, when the fraction of missing data is large, the KNN imputation gives much better accuracy. Comparing these results with the PRF, which is able to handle the missing data automatically, we find that the accuracy from the PRF is as good as that from the KNN imputation if the fraction of missing values is not too large (e.g. $<$50\%). Otherwise, we start to see that the KNN imputation gives a significantly better result.

\begin{table}
 \caption{The accuracy of the optimized RF and PRF models (trained with 45 features) when considering the test set with different portion of missing data and different way to impute them.}
 \label{tab:missfeatures}
 \begin{tabular}{cccc}
  \hline
   \% Missing Values in Each Feature & RF with Mean Impute & RF with KNN Impute & PRF  \\
  \hline
  5 & 0.731 & 0.737  & 0.738    \\
  10 & 0.721 & 0.736 & 0.737   \\
  15 & 0.703 & 0.733 & 0.735   \\
  50 & 0.368 & 0.724 & 0.661  \\
  \hline
 \end{tabular}
\end{table}

With the results described in Section~\ref{sec:bestml} and in this section, we can see that RF is one of the algorithms that provides the best accuracy. RF is also relatively fast to train and does not require feature scaling. It is also relatively easy to understand and interpret, and we can directly obtain the most important features that drive the classification results, which will be discussed later in Section~\ref{sec:bestfeature}. Therefore, in the remainder of Section~\ref{sec:results}, we will consider only RF as our ML model. 

\subsection{The Most Important Features}
\label{sec:bestfeature}

The importance of a feature (as implemented in \textit{scikit-learn}\footnote{\label{foot:feat_imp}\url{ https://scikit-learn.org/stable/modules/generated/sklearn.tree.DecisionTreeClassifier.html}}) in a decision tree is defined as the (normalized) weighted sum of the reduction of the impurity criterion brought by that feature on all nodes. The reduction in each node is weighted by the fraction of examples reaching that node. The importance of a feature in RF is the average of this value over all trees in the forest. We find that the five most important features are \textit{r-z}, \textit{r-i}, \textit{r-J}, \textit{r-H}, and \textit{g-z}. We use only these features, with re-optimization of the relevant parameters, to train a new RF model, and we receive an accuracy of 0.737, which is comparable to the best RF model that uses all 45 features (Table~\ref{tab:features}).

\begin{table}
 \caption{The accuracy of the optimized RF models trained with the different set of features. The first number is the accuracy of the exact prediction and the second number is the accuracy of the prediction including $\pm$1 sub-type error.}
 \label{tab:features}
 \begin{tabular}{lccccccc}
  \hline
    & All Features 
    & Best 5 Features 
    & \textit{ugriz} 
    & \textit{JHK\textsubscript{s}} 
    & \textit{JHK\textsubscript{s}}W1W2 
    & \textit{riJHK\textsubscript{s}} 
    & \textit{riJHK\textsubscript{s}}W1W2 \\
   \hline
    RF 
    & 0.738, 0.996  
    & 0.737, 0.994  
    & 0.733, 0.994  
    & 0.227, 0.492  
    & 0.377, 0.742  
    & 0.730, 0.993  
    & 0.731, 0.995  
    \\
   \hline
 \end{tabular}
\end{table}

\subsection{Features from Available Surveys}
\label{sec:surveysfeat}

In this section, we consider the cases where the available features are defined by available photometric surveys. We consider the following cases: (1) only SDSS bands, \textit{ugriz}; (2) only 2MASS bands \textit{JHK\textsubscript{s}}; and 3) 2MASS and AllWISE bands, \textit{JHK\textsubscript{s}}W1W2. These cases are selected based on the availability of photometric surveys rather than that of individual bandpasses. We specifically consider the case (3) because we want to see the accuracy received from using only the NIR bands without the optical bands. We do not consider pure AllWISE bands because only one feature (i.e. W1-W2) can be created. We also consider two additional cases: (4) \textit{riJHK\textsubscript{s}} and 5) \textit{riJHK\textsubscript{s}}W1W2. These cases are considered because one might want to use the \textit{ri} magnitudes from the Pan-STARRS1 (PS1) survey \citep{Chambers2016} when the magnitudes from the SDSS are unavailable. The \textit{ri} bands of the PS1 are similar to those of the SDSS \citep{Tonry2012}, so they could be used when the SDSS magnitudes are unavailable. 
     
For each of the cases described above, we train an RF model using only the available corresponding bands. We find that by using only the SDSS bands, we can achieve an accuracy that is only $\sim$1\% less than using all 45 bands. Meanwhile, using the 2MASS and AllWISE bands together gives $\sim$50\% less accuracy when compared to using only the SDSS bands. When using only the 2MASS bands, the accuracy falls to $\sim$20\%. However, every case can provide an accuracy higher than the true randomness (10\%). This indicates that photometric data can provide useful information for spectral classification. When allowing the prediction error up to $\pm$1 sub-type, the accuracy from using only the SDSS bands is $\sim$99\% and the accuracy from using only the NIR bands increases by a factor of 2 from the exact prediction. These results are presented in Table~\ref{tab:features}.

We also perform experiments similar to those in Section~\ref{sec:missingfeature}. That is, after training the model using only the bands available based on the available surveys, we assume that a fraction of the stars in the test set have some random missing values within those bands and the missing values are handled by KNN imputation. The results are shown in Table~\ref{tab:missfeatsurvey}. We can see that the imputation works very well in recovering random missing values within the measured bands.

\begin{table}
 \caption{The accuracy of the optimized RF models when considering different portion of missing data on the test set, which are handled by KNN imputation (k=20). The optimized RF models are trained using different set of features based on the photometric bands of the available surveys.}
 \label{tab:missfeatsurvey}
 \begin{tabular}{lcccccc}
  \hline
   \% Missing Values in Each Feature & All Features & \textit{ugriz} & \textit{JHK\textsubscript{s}} & \textit{JHK\textsubscript{s}}W1W2 & \textit{riJHK\textsubscript{s}} & \textit{riJHK\textsubscript{s}}W1W2 \\
  \hline
  no missing values & 0.738 & 0.733 & 0.227 & 0.377 & 0.730 & 0.731 \\
  5  & 0.737 & 0.732 & 0.226 & 0.377 & 0.728 & 0.730 \\
  10 & 0.736 & 0.731 & 0.224 & 0.377 & 0.728 & 0.730 \\
  15 & 0.733 & 0.731 & 0.223 & 0.377 & 0.727 & 0.730\\
  50 & 0.724 & 0.717 & 0.199 & 0.367 & 0.714 & 0.725 \\
  \hline
 \end{tabular}
\end{table}

Finally, we perform an additional experiment by using the RF model trained from all 45 features. We assume we have missing features on the test set based on the availability of photometric surveys (as describe above) and we impute those missing features using KNN imputation to make each example has all 45 features. We then predict the spectral sub-type of the example. We find that the accuracy in cases (1), (2), (3), (4), and (5) decreases by 0.001, 0.064, 0.043, 0.004, and 0.007, respectively, when compared to the results in Table~\ref{tab:features}. These results indicate that it is better to have several ML models, each trained with the features matching the bands available from photometric surveys, rather than training a ML model with all 45 features and then applying imputation for the missing measurement bands, especially in cases (2) and (3). In Section~\ref{sec:app}, we demonstrate an application of this study to the data set where the spectral sub-type is unavailable.

\subsection{Effect of Dust Extinction}
\label{sec:dust} 
\begin{figure}
	\includegraphics[width=0.9\textwidth]{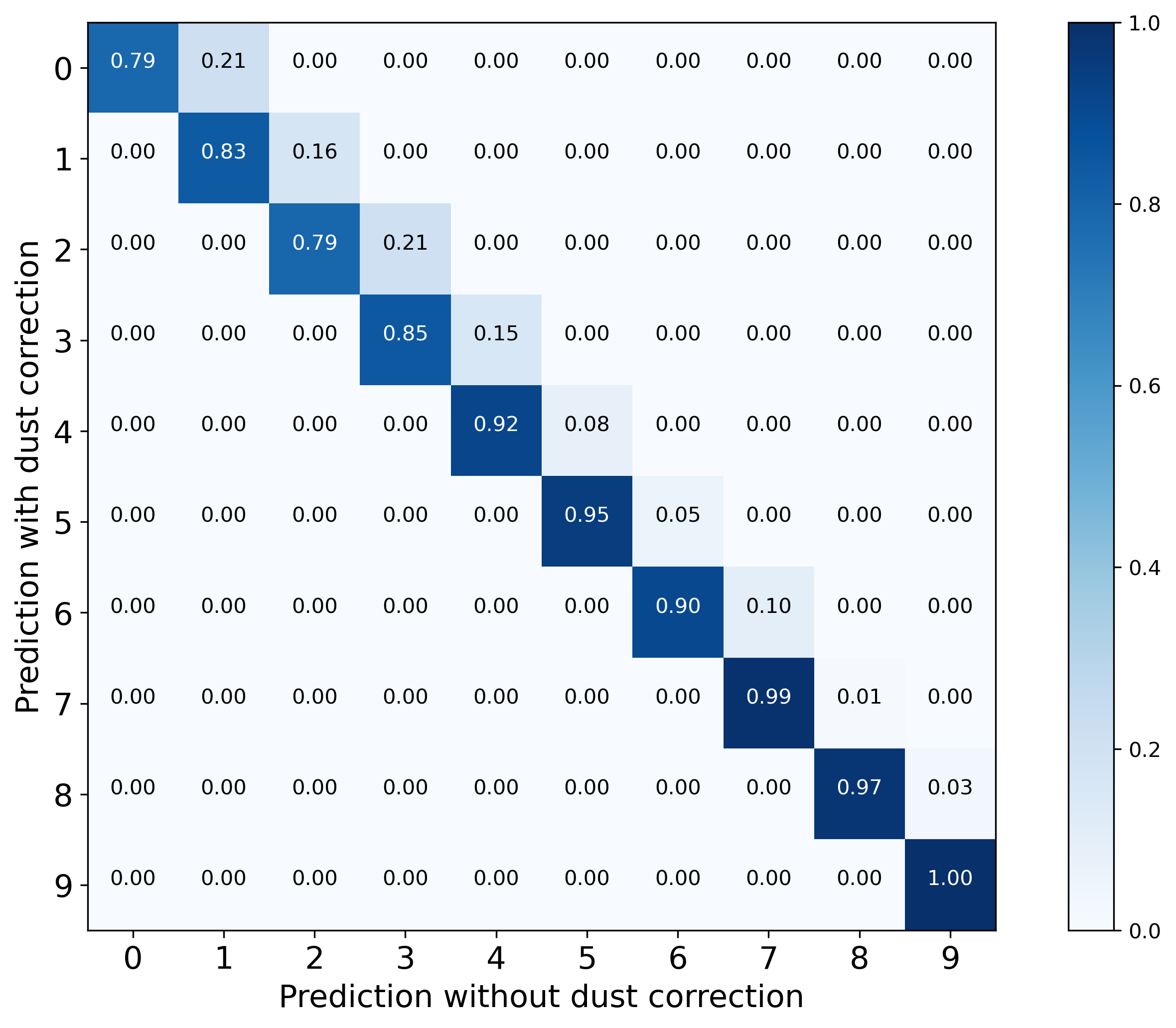}
    \caption{Modified confusion matrix comparing the predictions of examples with and without dust correction.}
    \label{fig:cfmcompare}
\end{figure}

In this section, we investigate the effect of the dust extinction on the spectral sub-type prediction accuracy. Specifically, we determine the change in the prediction accuracy if we use the model created from dust-corrected data, like those in the previous sections, to predict an example that is not corrected for dust reddening. We use the optimized RF model created from the 45 features with dust correction to predict spectral sub-types of examples in the test set where their features are not corrected for dust extinction. We obtain an overall accuracy of 0.714, which is lower by only 3\%. Figure~\ref{fig:cfmcompare} shows a modified confusion matrix where, instead of comparing the predicted spectral sub-types to the ground truth, the resulting predictions of the examples in the test set with and without dust correction are compared. We find that without dust correction, the predictions are shifted by at most +1 sub-type and the fraction of the examples that is shifted decreases with the increasing sub-type. These results are expected because without dust correction the stars appear redder and early-type stars distribute to larger distances, which are more affected by dust extinction.    

\begin{figure}
	\includegraphics[width=0.9\textwidth]{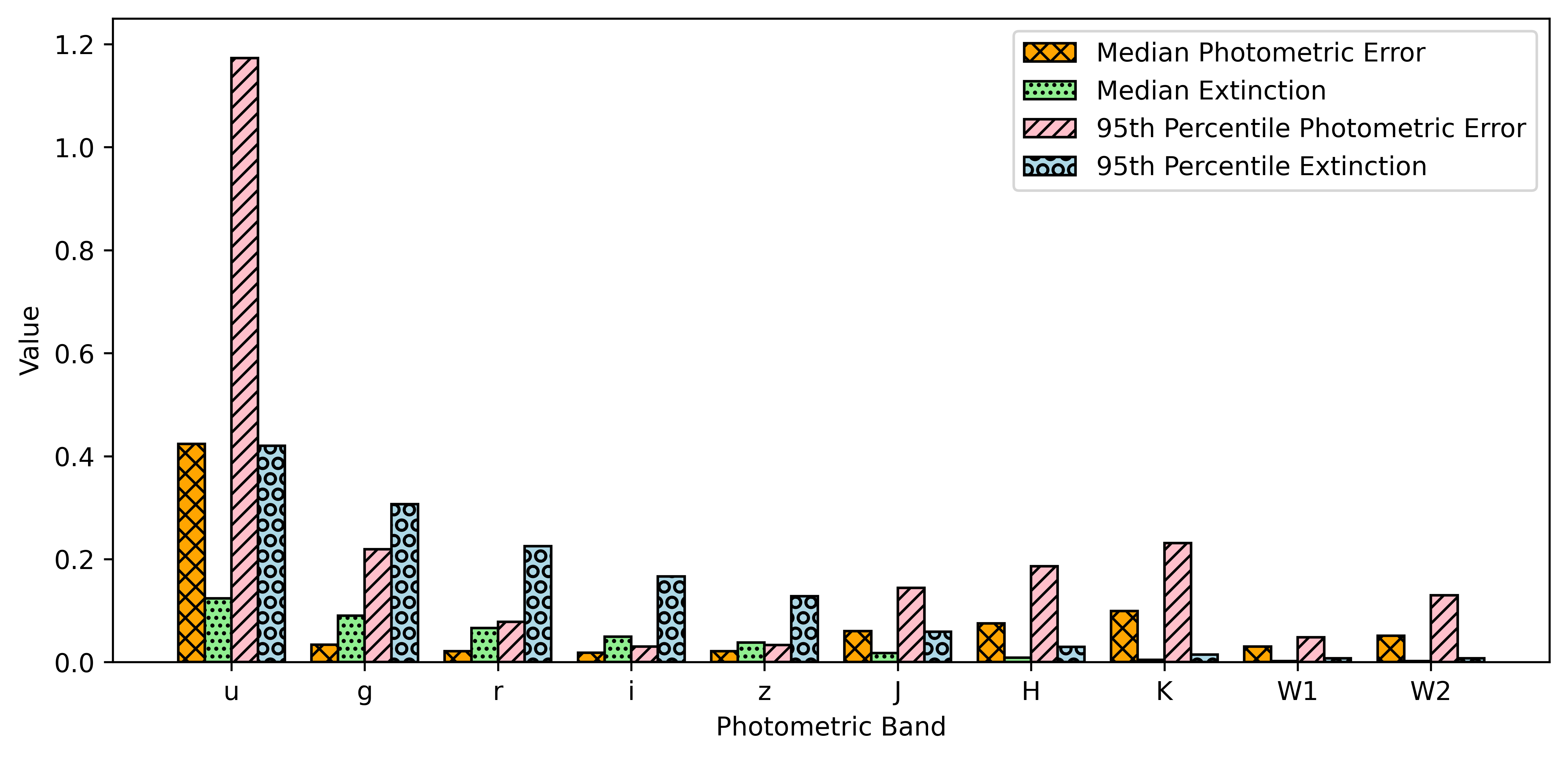}
    \caption{The median and 95th percentile values of the photometric error and dust extinction in each band for the stars that were selected in Section~\ref{sec:Data} and used to build the ML models.}
    \label{fig:photerrdust}
\end{figure}

Note that the stars that we use to train and test the models in this work are nearby, where 95\% of the stars locate within $\sim$1300 pc and their distribution can be determined from Table~\ref{tab:samplestat}. Figure~\ref{fig:photerrdust} illustrates the magnitudes of the extinction and photometric error in each band for the stars in this study. 
If a new star to be predicted has a similar or lower level of dust extinction and photometric error compared to these values, then the prediction accuracy should not be significantly worse. 

\section{Application}
\label{sec:app}
In this section, we use the optimized RF models that were created previously to predict spectral sub-types of stars in the Catalog of Cool Dwarf Targets for the Transiting Exoplanet Survey Satellite \citep{Muirhead2018}. We obtain these stars from the Exoplanet CTL v08.01 cross-matched with the TIC v8.1 catalog\footnote{\url{https://archive.stsci.edu/tess/tic_ctl.html}}, and only the entries with \texttt{cooldwarfs\_v8} flag are selected, resulting in 3,940,291 stars in the sample hereafter CDS8. These stars are then arranged into seven groups based on the criteria described below, and their spectral sub-types are predicted using the models built corresponding to those criteria with the accuracy reported in Table~\ref{tab:features}.

\textit{Group 0} includes all stars that are in the W11 catalog. The spectral sub-types of these stars are drawn from the W11 catalog and not predicted by our models. 

\textit{Group 1} includes the stars that have measured magnitudes available in all 10 bands, \textit{ugrizJHK\textsubscript{s}}W1W2. The predicted sub-types of these stars are 99.6\% accurate within 1 sub-type of the spectroscopic sub-types.

\textit{Group 2} includes the stars that have only the SDSS \textit{ugriz} magnitudes. We use the model constructed using only these magnitudes to predict the spectral sub-types and the prediction accuracy within one sub-type is 99.4\%.

\textit{Group 3} includes the stars that have \textit{JHK\textsubscript{s}}W1W2 magnitudes and \textit{ri} magnitudes from the Pan-STARRS1 (PS1) survey, which are used as the substituting magnitudes of the unavailable SDSS \textit{ri} magnitudes. We obtain the PS1 magnitudes by searching within 2 $\arcsec$ around a CDS8 star to find the closest PS1 star. It is possible for the closest star to actually be a different star. We estimate the percentage of the matching error by comparing the SDSS \textit{ri} magnitudes of the CDS8 star to the \textit{ri} magnitudes of the found PS1 star. There are around 800,000 matched stars that have the \textit{ri} magnitudes from both surveys for the comparison. We find that 79\% of the stars have the difference in the \textit{r} magnitudes and the difference in the \textit{i} magnitudes less than the 95th percentile of the magnitude errors in the corresponding bands of the stars that we used to train the models  (see Figure~\ref{fig:photerrdust}). In addition, 96\% of the stars have \textit{ri} magnitude differences less than the 95th percentile of the dust extinction values in the corresponding bands. We have learnt previously that this level of dust extinction can reduce the prediction accuracy by only $\sim$3\%. Therefore, we adopt the 4\% as the matching error, and the prediction accuracy within 1 sub-type is estimated from (99.5\%-3\%)x96\% $\sim$ 92.6\%. 

\textit{Group 4} includes of the stars that have \textit{JHK\textsubscript{s}} magnitudes and \textit{ri} magnitudes from the PS1. The process to match the CDS8 stars to the PS1 stars is the same as that of \textit{Group 3}, as well as the estimation of the matching error and prediction accuracy within 1 sub-type, which results in $\sim$92.4\%.

\textit{Group 5} includes the stars have only \textit{JHK\textsubscript{s}}W1W2 magnitudes. The prediction accuracy within 1 sub-type is 74.2\%.

The stars in \textit{Group 6} have only \textit{JHK\textsubscript{s}} magnitudes. The spectral sub-type prediction is 49.2\% accurate within 1 sub-type. 

Finally, \textit{Group 7} includes the stars that do not have any photometric information in the 10 bands. Therefore, we are unable to predict the spectral sub-type of these stars.

When creating features for the prediction, we also correct for dust extinction, as we did when building the models. The CDS8 catalog provides \textit{E(B-V)} values for the stars locating beyond 100 pc, and there are two sources of the \textit{E(B-V)}: Schlegel dust map and Pan-STARRS dust map. For the Schlegel dust map, we calculate the dust extinction using the extinction vector (\textit{R}) from Table 6 of \citet{Schlafly2011}, assuming \textit{R\textsubscript{V}} = 3.1, if R is available for the given band, and we do not correct for the extinction for the band without the \textit{R} value. We estimate the effect of correcting only the bands with available \textit{R} (leaving the other bands uncorrected), similar to what we did in Section~\ref{sec:dust}. We find that this affects the prediction accuracy by less than 1\%. For the Pan-STARRS, the extinction vector values that we use come from Table 1 of \citet{Green2018}. For the bands where the values are unavailable, we use the extrapolated values, such as those we adopted when training the models in Section~\ref{sec:Method}.     

Before we make a prediction for each star, we need to check whether all of the features of the star are in the value ranges of the features that we used to build the models. We do this because the ML models are unable to predict the result by extrapolation. Because the CDS8 also contains late-K dwarfs, we need to exclude them from the prediction. Table~\ref{tab:cds8} summarizes the total number of stars in each group and the number of stars that are not classified by our ML models.

\begin{table}
 \caption{The number of stars in the CDS8 arranged into different groups based on the criteria explained in the text, the number of those stars that are not predicted by our ML models because their features are not in the value ranges of the corresponding features used to train the models, and the prediction accuracy corresponding to each group.}
 \label{tab:cds8}
 \begin{tabular}{cccc}
  \hline
  Group & All & Not Predicted & Expected Accuracy  \\
   & [Counts] & [Counts] & within $\pm$1 Sub-Type [\%] \\
  \hline
  0 & 6998 & 6998 & N/A \\ 
  1 & 795198 & 59945 & 99.6 \\ 
  2 & 18738 & 5095 & 99.4 \\ 
  3 & 1979393 & 124762 & 92.6 \\
  4 & 110649 & 14533 & 92.4 \\
  5 & 934955 & 24243 & 74.2\\
  6 & 94337 & 180 & 49.2 \\
  7 & 23 & 23 & N/A \\
  \hline
  All & 3940291 & 235779 \\
  \hline
 \end{tabular}
\end{table}

The results from the predictions in this section are available for downloading at our GitHub repository: \url{https://github.com/sirinrats/ml-md-class/}.

\section{Available Models}
\label{sec:models}

We provide the optimized RF models that we created and used in Section~\ref{sec:app} for public use. We briefly describe the models again here for reference. The models were trained with different number of features, as summarized below:

\textit{Model 1} was trained with colors derived from all 10 bands, \textit{ugrizJHK\textsubscript{s}}W1W2.

\textit{Model 2} was trained with colors derived from \textit{ugriz} bands.

\textit{Model 3} was trained with colors derived from \textit{riJHK\textsubscript{s}}W1W2 bands.

\textit{Model 4} was trained with colors derived from \textit{riJHK\textsubscript{s}} bands.

\textit{Model 5} was trained with colors derived from \textit{JHK\textsubscript{s}}W1W2 bands.

\textit{Model 6} was trained with colors derived from \textit{JHK\textsubscript{s}} bands.

All of the models listed above were trained without feature scaling and resampling. The hyperparameters were optimized through 10-fold cross-validation. For each model, when the required features are missing, the missing features will be imputed using the KNN imputation. The user can decide which model(s) to be used, based on the availability of the features of the stars to be predicted. For example, if W1 and W2 are missing, then the user can use the Model 1 where the features related to W1 and W2 will be imputed by the model, with the caveat that the expected accuracy will drop (refer to Table~\ref{tab:missfeatures}) or the user can use \textit{Model 2} with the expected accuracy as reported in Table~\ref{tab:features}. 

The models listed above were trained using the examples with the spectral sub-type distribution inherited from the W11 sample (Table~\ref{tab:samplestat}). If the user thinks that the distribution of the sample to be classified is different from the W11 sample, then they might want to use the models trained with the sample that has the sub-type uniformly distributed. Therefore, we also provide the corresponding models with Random Undersampling applied.

The models, corresponding confusion matrices, and reported accuracy are available at our GitHub repository. An example snippet code to run the model is also provided.   

\section{Summary}
\label{sec:summary}
We examined how well machine-learning models could classify the different spectral sub-types of M dwarfs by utilizing only their photometric magnitudes in the optical and NIR bands. The models that we considered were KNN, RF, PRF, and MLP, and the magnitudes that we used were from the SDSS-\textit{ugriz} bands, 2MASS-\textit{JHK\textsubscript{s}} bands, and AllWISE-W1W2 bands. The findings indicate that all of the models that we tested were capable of correctly classifying over 70\% of M dwarfs into the corresponding sub-type, as determined by spectroscopic analysis. Additionally, the models were able to classify more than 99\% of the M dwarfs into sub-types within a margin of error of $\pm$1 sub-type when compared to the results of spectroscopic classification. The most useful wave bands for the classification were \textit{r}, \textit{z}, \textit{i}, \textit{J}, \textit{H}, and \textit{g} bands. Given that the stars examined in this study were relatively close to the Earth ($\lesssim$ 1300 pc), the effect of dust extinction was small, degrading the classification accuracy by only 3\%. 

Resampling techniques such as SMOTE or Random Undersampling, which adjust the number of examples in each class to be evenly distributed, could significantly improve the predictive performance of minority classes. However, with our test set, which has an uneven distribution inherited from the SDSS survey, the overall accuracy decreased by 2\%. 

We employed the RF models, which demonstrated the highest level of performance among the machine-learning models that we evaluated, and classified the spectral sub-type (M0V-M9V) of 94\% of stars in the ~\citet{Muirhead2018} catalog. Furthermore, we have made the models available for public use via our GitHub repository.

It should be noted that the concept of model performance, like accuracy, only makes sense when considering a sample of data. When using an ML model to predict a single target, the result can be either correct or incorrect. Therefore, the provided ML models should not be used for the study of an individual star but rather used in the study of a large sample.  

This research aims to contribute to the development of new and innovative methods for analyzing large astronomical data sets that are difficult to manage using traditional methods. It provides insight into the level of accuracy that can be achieved under certain resources and constraints. This research could prove particularly beneficial for upcoming surveys such as the Large Synoptic Survey Telescope (LSST; ~\citet{Ivezic2019}), which is expected to generate an enormous amount of data that would be impossible to analyze using conventional techniques.

\section{Data Access} \label{sec:pubdata}

The public data described here may be obtained from the following websites.

\textit{SDSS DR7 M Dwarfs (J/AJ/141/97/catalog) :} 
\url{https://vizier.cds.unistra.fr/}

\textit{2MASS :}
\url{https://irsa.ipac.caltech.edu/Missions/2mass.html}

\textit{AllWISE :} 
\url{https://wise2.ipac.caltech.edu/docs/release/allwise/}

\textit{Gaia DR3 :} 
\url{https://gea.esac.esa.int/archive/}

\textit{Gaia eDR3 distances:} 
\url{https://gea.esac.esa.int/archive/}

\textit{Wide Binaries from Gaia eDR3 :} 
\url{https://zenodo.org/record/4435257#.Y_QyLuxBzvU}

\textit{TESS :} 
\url{https://archive.stsci.edu/tess/tic_ctl.html}

\textit{Pan-STARRS1 :} 
\url{https://catalogs.mast.stsci.edu/panstarrs/}

\section{Acknowledgments}
The authors would like to thank Dr. Apichart Hortiangtham for insights, suggestions, and fruitful discussions on machine learning methods. The authors would like to thank the anonymous referees for their suggestions on how to enhance this manuscript. 

This research was funded by the National Astronomical Research Institute of Thailand (Public Organization) (NARIT) and Thailand Science Research and Innovation (TSRI). The authors extend our sincere appreciation to NARIT and TSRI for their support.

This research made use of the NASA Astrophysics Data System (ADS) Bibliographic Services, the SIMBAD database ~\citep{Wenger2000}, the cross-match service provided by CDS Strasbourg, \texttt{scikit-learn} ~\citep{Pedregosa2011}, \texttt{NumPy} ~\citep{harris2020array}, \texttt{pandas} ~\citep{mckinney-proc-scipy-2010}, and \texttt{Keras} ~\citep{chollet2015keras}.  

This work made extensive use of the Sloan Digital Sky Survey (SDSS) data. Funding for the SDSS and SDSS-II has been provided by the Alfred P. Sloan Foundation, the Participating Institutions, the National Science Foundation, the U.S. Department of Energy, the National Aeronautics and Space Administration, the Japanese Monbukagakusho, and the Max Planck Society, and the Higher Education Funding Council for England. The SDSS is managed by the Astrophysical Research Consortium (ARC) for the Participating Institutions. The Participating Institutions are the American Museum of Natural History, Astrophysical Institute Potsdam, University of Basel, University of Cambridge, Case Western Reserve University, The University of Chicago, Drexel University, Fermilab, the Institute for Advanced Study, the Japan Participation Group, The Johns Hopkins University, the Joint Institute for Nuclear Astrophysics, the Kavli Institute for Particle Astrophysics and Cosmology, the Korean Scientist Group, the Chinese Academy of Sciences (LAMOST), Los Alamos National Laboratory, the Max-Planck-Institute for Astronomy (MPIA), the Max-Planck-Institute for Astrophysics (MPA), New Mexico State University, Ohio State University, University of Pittsburgh, University of Portsmouth, Princeton University, the United States Naval Observatory, and the University of Washington. The SDSS Web site is \url{http://www.sdss.org/}.

This publication made use of data products from the Two Micron All Sky Survey (2MASS), which is a joint project of the University of Massachusetts and the Infrared Processing and Analysis Center/California Institute of Technology, funded by the National Aeronautics and Space Administration and the National Science Foundation.

This publication made use of data products from the Wide-field Infrared Survey Explorer (\textit{WISE}), which is a joint project of the University of California, Los Angeles, and the Jet Propulsion Laboratory/California Institute of Technology, funded by the National Aeronautics and Space Administration.

This paper includes data collected with the \textit{TESS} mission, obtained from the MAST data archive at the Space Telescope Science Institute (STScI). Funding for the \textit{TESS} mission is provided by the NASA Explorer Program. STScI is operated by the Association of Universities for Research in Astronomy, Inc., under NASA contract NAS 5–26555.

This publication made use of data products from the Pan-STARRS1 (PS1) Surveys. The PS1 and the PS1 public science archive have been made possible through contributions by the Institute for Astronomy, the University of Hawaii, the Pan-STARRS Project Office, the Max-Planck Society and its participating institutes, the Max Planck Institute for Astronomy, Heidelberg and the Max Planck Institute for Extraterrestrial Physics, Garching, The Johns Hopkins University, Durham University, the University of Edinburgh, the Queen's University Belfast, the Harvard-Smithsonian Center for Astrophysics, the Las Cumbres Observatory Global Telescope Network Incorporated, the National Central University of Taiwan, the Space Telescope Science Institute, the National Aeronautics and Space Administration under Grant No. NNX08AR22G issued through the Planetary Science Division of the NASA Science Mission Directorate, the National Science Foundation Grant No. AST–1238877, the University of Maryland, Eotvos Lorand University (ELTE), the Los Alamos National Laboratory, and the Gordon and Betty Moore Foundation.

This work made use of data from the European Space Agency (ESA) mission
{\it Gaia} (\url{https://www.cosmos.esa.int/gaia}), processed by the {\it Gaia}
Data Processing and Analysis Consortium (DPAC,
\url{https://www.cosmos.esa.int/web/gaia/dpac/consortium}). Funding for the DPAC
has been provided by national institutions, in particular the institutions
participating in the {\it Gaia} Multilateral Agreement.

\bibliography{PASP_MDW}{}
\bibliographystyle{aasjournal}



\end{document}